\documentclass[10pt,twocolumn]{article} 

\usepackage{ol2_no_copyright}

\usepackage{textcomp}
\usepackage{epstopdf}

\usepackage{graphicx, latexsym, verbatim}
\usepackage{graphics}
\usepackage{amssymb, amsmath}
\usepackage[ansinew]{inputenc}
\newcommand{\ud}{\mathrm{d}}
\newcommand{\mr}{\mathbf{r}}

\newcommand{\mk}{\mathbf{k}}

\newcommand{\mf}{\mathbf{f}}
\newcommand{\mkt}{\tilde{\mathbf{k}}}
\newcommand{\mft}{\tilde{\mathbf{f}}}
\newcommand{\tlo}{\tilde{\omega}}
\newcommand{\ft}{\tilde{\text{f}}}

\newcommand{\me}{\mathbf{e}}

\newcommand{\rev}[1]{{#1}}
\newcommand{\comm}[1]{{#1}}

\begin{document}
\twocolumn[ 

\bibliographystyle{unsrt}

\title{Calculation, normalization and perturbation of quasinormal modes in coupled cavity-waveguide systems}

\author{Philip Tr{\o}st Kristensen,$^{1,2*}$ Jakob Rosenkrantz de Lasson,$^{1}$ and Niels Gregersen$^{1}$}

\address{
$^1$DTU Fotonik, Technical University of Denmark, DK-2800 Kgs. Lyngby, Denmark
\\
$^2$Institut f{\"u}r Physik, Humboldt Universit{\"a}t zu Berlin, D-12489 Berlin, Germany\\
$^*$Corresponding author: philip.kristensen@physik.hu-berlin.de}

\begin{abstract}
We show how one can use a non-local boundary condition, which is compatible with standard frequency domain methods, for numerical calculation of quasinormal modes in optical cavities coupled to waveguides. In addition, we extend the definition of the quasinormal mode norm by use of the theory of divergent series to provide a framework for modeling of optical phenomena in such coupled cavity-waveguide systems. As an example, we apply the framework to study perturbative changes in the resonance frequency and $Q$ value of a photonic crystal cavity coupled to a defect waveguide. 
\end{abstract}


]


\maketitle

Quasinormal modes (QNMs)~\cite{Ching_RevModPhys_70_1545_1998, Ching_1996, Kristensen_ACSphot_1_2_2014} provide a natural framework for modeling light propagation and light-matter interaction in resonant electromagnetic material systems, such as optical cavities~\cite{Lai_PRA_41_5187_1990, Leung_PRA_49_3057_1994, Leung_JOSAB_13_805_1996, Lee_JOSAB_16_1409_1999, Lee_JOSAB_16_1418_1999, Kristensen_OL_37_1649_2012, Maes_OE_21_6794_2013} and plasmonic nanoparticles~\cite{Sauvan_PRL_110_237401_2013, deLasson_JOSAB_30_1996_2013, Bai_OE_21_27371_2013, Ge_arXiv_1312.2939_2013, Ge_OL_39_4235_2014}. 
The QNMs of localized electromagnetic resonators in an otherwise homogeneous medium may be calculated as eigenmodes of the source-free Maxwell equations augmented with the Silver-Müller radiation condition~\cite{Kristensen_ACSphot_1_2_2014}. The radiation condition admits only fields that propagate away from the resonator at large distances. This, in turn, leads to a discrete spectrum of complex resonance frequencies $\tlo_\mu =\omega_\mu - \text{i}\gamma_\mu$ from which the $Q$ value may be calculated directly as $Q=\omega_\mu/2\gamma_\mu$~\cite{Lalanne_LaserPhotRev_2_514_2008}. In numerical calculations, the radiation condition is often modeled using perfectly matched layers (PMLs)~\cite{Berenger_JCP_114__185_1994} in either frequency- or time-domain calculations, but alternatives are also available, for example in the form of integral equations
~\cite{Kristensen_OL_37_1649_2012, deLasson_JOSAB_30_1996_2013}. 

\comm{In many technologically relevant applications, light coupling to and from the cavity is controlled by waveguides, such as fibers or line defects in photonic crystal (PC)~\cite{Joannopoulos2008} circuits. In such} coupled cavity-waveguide systems, the coupling to the waveguides represents by far the largest decay channel for light in the cavity; indeed, any decay \comm{through} other channels than the waveguide often represents unwanted and detrimental losses and is typically minimized through careful engineering. \comm{Calculation of QNMs in these coupled systems is non-trivial because of the need for a suitable radiation condition. In particular, the Silver-Müller radiation condition applies only to problems with a homogeneous background material, and if the waveguides have discrete translational symmetry --- as in PCs, for example --- one cannot use PMLs to avoid reflections from the calculation domain boundary. One suitable calculation method for such problems is} the Fourier Modal Method (FMM)
~\cite{Moharam_JOSA_71_811_1981, Noponen_JOSAA_11_2494_1994,Lecamp_OE_15_11048_2007}, 
which can easily incorporate the radiative dissipation of a QNM leaking from the resonator through the waveguide\comm{~\cite{deLasson_arXiv_1405_2595_2014}}. 

\begin{figure}[htb]
\includegraphics[width=\columnwidth]{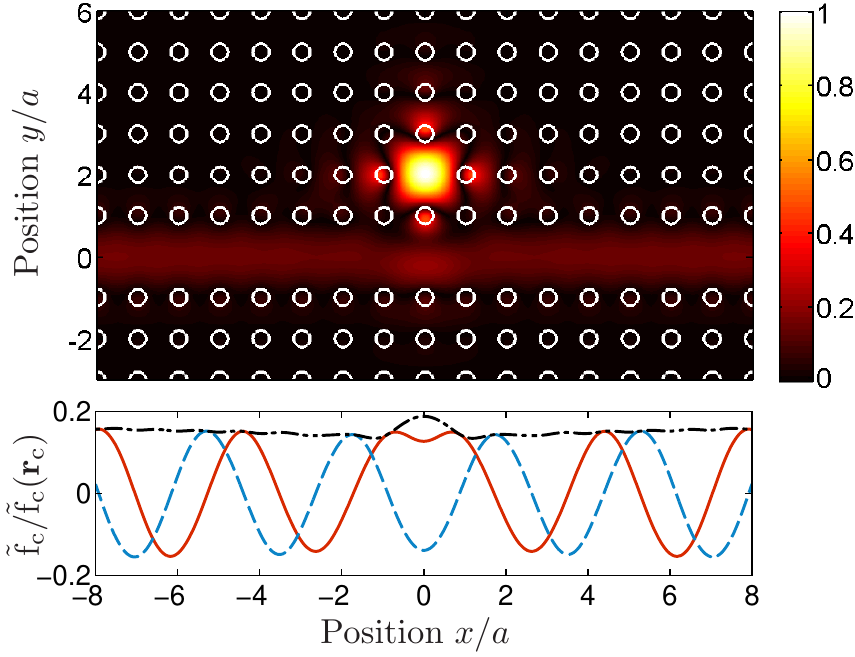}
\caption{\label{Fig:side_coupled_2D_quadratic_rods_oneBlock_cavityMode}Top: Absolute value of the QNM $|\ft_\text{c}(\mr)|$ in \comm{a cavity side-coupled to an infinite waveguide in a PC with lattice constant $a$}. \comm{The QNM has a complex resonance frequency of $\tlo_\text{c}a/2\pi\text{c}=0.3969 - 0.0014\text{i}$, corresponding to $Q=146$}. Bottom: Real part (red full), imaginary part (blue dashed) and absolute value (black dashed-dotted) of the QNM along the line $y=0$ in the center of the waveguide. In both panels, the QNM is scaled to unity at $\mr=\mr_\text{c}$ in the center of the cavity. 
} \end{figure}

\comm{In this Letter we present an alternative to the FMM based on a non-local boundary condition, which can be used with standard frequency domain methods to calculate QNMs in coupled cavity-waveguide systems.} Figure \ref{Fig:side_coupled_2D_quadratic_rods_oneBlock_cavityMode} shows the \comm{QNM of a side-coupled cavity in a PC with lattice constant $a$} made from high-index cylinders \comm{($\epsilon_\text{rods}=8.9$)} in air. This QNM --- which is identical to the QNM that was calculated with the FMM in Ref.~\cite{deLasson_arXiv_1405_2595_2014} --- was calculated using the finite element method (FEM) 
with the special boundary condition to properly model the waveguide radiation. 

The leaky nature of the QNMs leads to an exponential divergence at large distances, which means that they cannot be normalized in the same way as the so-called normal modes calculated using Dirichlet (or periodic) boundary conditions. Instead, the QNMs may be normalized by an alternative prescription that explicitly compensates for the divergence. This normalization was first derived for spherically symmetric material systems in Refs.~\cite{Lai_PRA_41_5187_1990, Leung_PRA_49_3057_1994, Leung_JOSAB_13_805_1996, Lee_JOSAB_16_1409_1999, Lee_JOSAB_16_1418_1999} 
and applied to general optical cavities in Ref.~\cite{Kristensen_OL_37_1649_2012}, which also introduced a generalized effective mode volume \comm{for} leaky optical cavities. Using the Lorentz reciprocity theorem, Sauvan \emph{et al.}~\cite{Sauvan_PRL_110_237401_2013} later derived an alternative formulation of the norm in an elegant and very transparent way. When the Silver-Müller radiation condition applies, the two formulations of the norm can be shown to be identical~\cite{Ge_arXiv_1312.2939_2013}. As noted above, this is not the case for cavities coupled to PC waveguides. Whereas the norm in Refs.\cite{Lai_PRA_41_5187_1990, Leung_PRA_49_3057_1994, Leung_JOSAB_13_805_1996, Lee_JOSAB_16_1409_1999, Lee_JOSAB_16_1418_1999} rely explicitly on the Silver-Müller condition, the formulation in Ref.~\cite{Sauvan_PRL_110_237401_2013} does not. 
\comm{Therefore, one could expect this formulation to be general enough to also handle 
QNMs leaking through waveguides but, as we show below, a direct application of the norm in Ref.~\cite{Sauvan_PRL_110_237401_2013} leads to a divergent integrand. 
As a remedy,} we show in this Letter how one can use the theory of divergent series
~\cite{Hardy_1949} to assign a finite value to the integral, and we use the normalization in an example problem of perturbation theory \comm{for the PC cavity-waveguide system in Fig.~\ref{Fig:side_coupled_2D_quadratic_rods_oneBlock_cavityMode}}.

We consider systems of cavities coupled to waveguides which are defined in general by a periodic relative permittivity distribution $\epsilon_\text{r}(\mr)$ for which $\epsilon_\text{r}(\mr+\mathbf{R})=\epsilon_\text{r}(\mr)$, where $\mathbf{R}$ is a lattice vector in the direction of the waveguide. In addition, we limit the analysis to non-magnetic, isotropic and dispersionless materials, and to cavities coupled to waveguides with a single \comm{band of defect waveguide modes} in the frequency range of interest. This is the case, for example, for the material system in Fig.~\ref{Fig:side_coupled_2D_quadratic_rods_oneBlock_cavityMode}\comm{, cf. chapter 5 of Ref.~\cite{Joannopoulos2008}}. \comm{We focus on electric field QNMs defined as} solutions to the wave equation
\begin{align}
\nabla\times\nabla\times \mft_\mu(\mr)-\Big(\frac{\tlo_\mu}{\text{c}}\Big)^2\epsilon_\text{r}(\mr)\mft_\mu(\mr) = 0,
\label{Eq:HelmholtzEq}
\end{align}
where $\text{c}$ is the speed of light, subject to a suitable radiation condition describing the light-propagation through the waveguide away from the cavity. In the periodic waveguides, Bloch-Floquet theory\comm{~\cite{Joannopoulos2008}} ensures that the solutions to the wave equation may be written as 
\begin{align}
\mf_\mk(\mr) = \exp(\text{i}\mk\cdot\mr)\mathbf{u}_\mk(\mr),
\label{Eq:Blochform}
\end{align}
in which $\mk$ is the wave vector and $\mathbf{u}_\mk(\mr+\mathbf{R}) = \mathbf{u}_\mk(\mr)$. At sufficient distances from the cavity, the leaky cavity mode may be written in the form of a single Bloch mode as in Eq.~(\ref{Eq:Blochform}) but with a complex wave vector $\mkt_\mu$ corresponding to the resonance frequency $\tlo_\mu$ and pointing in the direction of the waveguide away from the cavity. We assume the calculation domain boundary $\partial V$ to be a plane perpendicular to the waveguide direction (a line in two dimensions) with outward pointing unit vector $\mathbf{n}$. In this case, Eq.~(\ref{Eq:Blochform}) can be used to define the waveguide radiation condition as a nonlocal boundary condition of the form
\begin{align}
\mft_{\mkt}(\mr)\Big\vert_{\mr\in\partial V} = \exp(i\mkt\cdot\mathbf{n}a) \mft_{\mkt}(\mr-\mathbf{n}a).  
\label{Eq:waveguideRadiationCondition}
\end{align}
This non-local boundary condition is similar to the one introduced in Ref.~\cite{Hu_OE_29_1356_2012}, but the application is different in that there is no incoming field; the QNMs are defined as the solutions to the wave equation with no sources. Using Eq.~(\ref{Eq:waveguideRadiationCondition}), one can calculate QNMs of coupled cavity-waveguide systems using standard frequency domain methods. For most problems of practical interest, however, the dispersion of the waveguide modes has no closed form expression, and this complicates the use of Eq.~(\ref{Eq:waveguideRadiationCondition}) in QNM calculations for which the complex resonance frequency is the eigenvalue of interest. In practice, therefore, \comm{we calculate the complex wave vector $\mkt_\mu$} using a Taylor expansion approximation of the dispersion along the real frequency axis, which may then be readily extended to the complex frequency plane by analytic continuation. 

\begin{figure}[htb]
\includegraphics[width=\columnwidth]{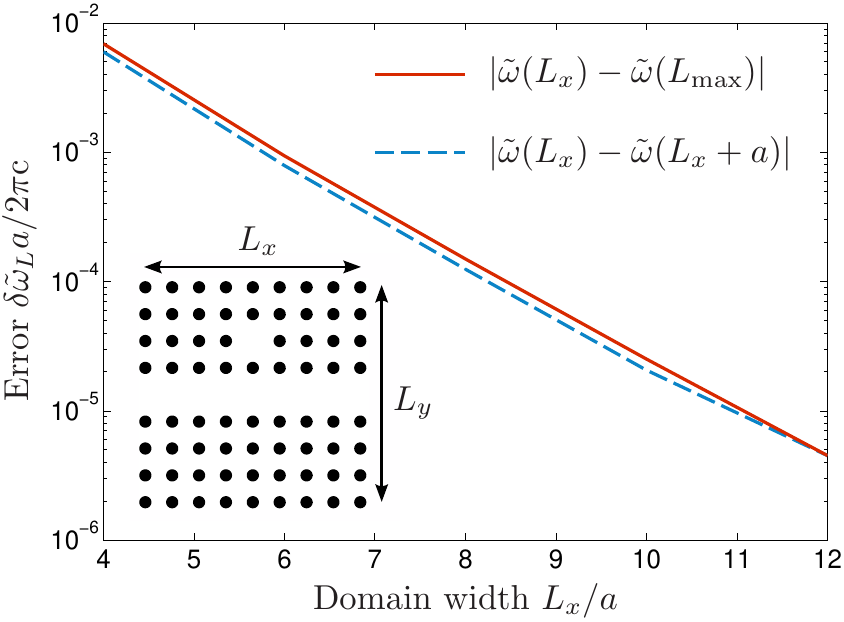}
\caption{\label{Fig:side_coupled_2D_quadratic_rods_oneBlock_freqDiff_vs_domainSize}\comm{Change in }resonance frequency with increasing calculation domain width $L_x$ \comm{as defined in the inset}. The red solid curve shows the difference to a reference calculation with fixed domain size $L_\text{max}=18a$, and the blue dashed curve shows the difference between results using successively larger domain sizes.}
\end{figure}

As with other frequency domain calculations of QNMs, the radiation condition results in a non-linear problem because it depends on the complex resonance frequency. 
Therefore, one must calculate the QNMs by an iterative procedure in which a fixed frequency $\tlo_\text{guess}$ is used to set up the equation system which is then subsequently solved to find the frequency closest to $\tlo_\text{guess}$. The iteration continues until the difference $\Delta\tlo$ is less than some prescribed tolerance $\Delta\tlo_\text{max}$. All calculations in this Letter were performed using FEM (Comsol Multiphysics 4.3a), and the iterations were performed with an absolute tolerance of $\Delta\tlo_\text{max}a/\text{c}=10^{-5}$. As for discretization, the number of elements was kept sufficiently high so as not to influence the results to the quoted number of digits. The analytic continuation of the dispersion curve into the complex \comm{frequency} plane was based on a fourth order polynomial fit to the real dispersion data around the point $\tlo_\text{R}a/2\pi\text{c}=0.395$. Comparing to the corresponding 5th order expansion reveals a difference $\Delta \tilde{k}a/2\pi<10^{-6}$ over an interval $|\tlo-\tlo_\text{R}|a/2\pi\text{c}<0.01$. This interval is more than an order of magnitude larger than the imaginary part of the calculated resonance frequency, wherefore we expect the error in the complex dispersion to be negligible for the calculations presented here. 
Finally, an important source of error in the calculations is linked to the size of the calculation domain and can be easily described in the language of the FMM~\cite{deLasson_arXiv_1405_2595_2014}. The FMM approach is based on an expansion using the full set of Bloch modes in the system --- with either propagating, growing or decaying characteristics. In the waveguides, the field is described by several decaying Bloch modes as well as a single growing Bloch mode~\cite{deLasson_arXiv_1405_2595_2014}, and Eq.~(\ref{Eq:waveguideRadiationCondition}) thus requires all purely decaying Bloch mode components of the QNM to be negligibly small at the calculation domain boundary. Figure~\ref{Fig:side_coupled_2D_quadratic_rods_oneBlock_freqDiff_vs_domainSize} illustrates the convergence of the QNM resonance frequency as the calculation domain size $L_x$ is increased. Specifically, it shows the error $\delta\tlo_L=|\tlo(L_x)-\tlo(L_\text{max})|$ when comparing to a fixed domain size $L_\text{max}=18a$ as well as the error $\delta\tlo_L=|\tlo(L_x)-\tlo(L_x+a)|$ when comparing the results obtained using successively larger domain sizes. In both cases, we used a fixed calculation domain height of $L_y=18a$. \comm{Because of the exponential decay of the QNM along the $y$-direction, this was found to be large enough to not influence the results to the quoted number of digits.} The error in the resonance frequency decreases exponentially with the domain size, consistent with the behavior of the non-propagating Bloch mode components of the QNM, which decay exponentially into the waveguide.

The procedure outlined above was applied to calculate the QNM in Fig.~\ref{Fig:side_coupled_2D_quadratic_rods_oneBlock_cavityMode}, which we denote by $\mu=\text{c}$ \comm{and which has only an out-of-plane component, $\mft_\text{c}(\mr)=\ft_\text{c}(\mr)\me_\text{z}$.} The QNM has a complex resonance frequency of $\tlo_\text{c}a/2\pi\text{c}=0.3969 - 0.0014\text{i}$, corresponding to $Q=146$. From analytic continuation of the dispersion we find that the magnitude of the complex wave vector in the direction of the waveguide is \comm{$\tilde{k}_\text{c}a/2\pi =  0.2837 - 0.0026\text{i}$}. 

Having calculated the QNM, we next turn to the question of normalization. Following Ref.~\cite{Sauvan_PRL_110_237401_2013}, we start by writing the norm of the QNM $\mft_\mu$ as
\begin{align}
\langle\langle\mft_\mu|\mft_\mu\rangle\rangle &= \frac{1}{2}\int_V\epsilon_\text{r}(\mr)\mft_\mu(\mr)\cdot\mft_\mu(\mr)\nonumber \\
 &\quad+ \Big(\frac{\tlo_\mu}{\text{c}}\Big)^2 \big(\nabla\times\mft_\mu(\mr)\big) \cdot\big(\nabla\times\mft_\mu(\mr)\big) \ud\mr,
\label{Eq:SauvanNorm}
\end{align}
where 
the integral is formally over the entire volume of space (area in two dimensions), \comm{and we have used the source-free Maxwell curl equation to express the integrand in terms of $\mft_\mu$ only}. When the QNM leaks into homogeneous surroundings, the integrand in Eq.~(\ref{Eq:SauvanNorm}) tends to zero at large distances from the cavity sufficiently fast to render the integral convergent. When the QNM leaks through a periodic waveguide this is not necessarily so; for the QNM in Fig.~\ref{Fig:side_coupled_2D_quadratic_rods_oneBlock_cavityMode} we find that in the waveguide, the integrand oscillates around zero and increases (exponentially) in magnitude as a function of the distance from \comm{the} cavity. The QNM in Fig.~\ref{Fig:side_coupled_2D_quadratic_rods_oneBlock_cavityMode} is clearly symmetric wherefore we limit the below calculations to $x>0$, but we note that the procedure generalizes immediately to non-symmetric material systems. Moreover, we split the $x$-integration into two parts at $x=x_0$ which is chosen sufficiently large that for $x>x_0$ the QNM in the waveguide is well described by Eq.~(\ref{Eq:Blochform}). For $x<x_0$, the integrand is well-behaved, and the integral can easily be evaluated. 
For the problematic part over the semi-infinite region $x>x_0$, we rewrite the integral as
\begin{align}
I_{x>x_0}&= I_a(x_0)\sum_{m=0}^\infty \text{e}^{2\text{i}\tilde{k}_\mu ma},
\label{Eq:divergentGeometric}
\end{align}
where 
\begin{align}
I_a(x_0) &=  \frac{1}{2}\int_\perp\int_{x_0}^{x_0+a}\epsilon_\text{r}(\mr)\mft_\mu(\mr)\cdot\mft_\mu(\mr)\nonumber\\ 
&\quad+ \Big(\frac{\tlo_\mu}{\text{c}}\Big)^2\big(\nabla\times\mft_\mu(\mr)\big) \cdot\big(\nabla\times\mft_\mu(\mr)\big) \ud x\ud\mr_\perp,
\end{align}
in which $\perp$ denotes the dimension(s) perpendicular to the $x$-axis (area in three dimensions and line in two dimensions). The sum in Eq.~(\ref{Eq:divergentGeometric}) is a geometric series of the form $\sum_m b^m$, 
which is formally divergent for $|b|\geq1$. \comm{The cavity mode in Fig.~\ref{Fig:side_coupled_2D_quadratic_rods_oneBlock_cavityMode} is an example of such a divergent case, for which we find that $\exp(2\text{i}\tilde{k}_\text{c}a)=-0.9421 - 0.4243\text{i}$. From Eq.~(\ref{Eq:divergentGeometric}), this can now be seen as the origin of the oscillating and divergent behavior of the integrand. Although the series is formally divergent, it is possible to assign a finite value to the right hand side of Eq.~(\ref{Eq:divergentGeometric}) using} Borel summation (for Re$\{b\}<1$) or Lindel{\"o}f or Mittag-Leffler summation (for $b\in\mathbb{C}\backslash[1,\infty[$)~\cite{Hardy_1949} , in which cases the sum evaluates to the same expression as in the case of \comm{$|b|<1$}, and one thus finds
\begin{align}
I_{x>x_0} = \frac{I_a}{1-\text{e}^{2\text{i}\tilde{k}_\mu a}}.
\label{Eq:I_Lambda}
\end{align}
Using a calculation domain size of $L_x=18a$ we choose $x_0=8a$ and scale the QNM to unity in the cavity center $\mr_\text{c}$, in which case the generalized effective mode volume~\cite{Kristensen_OL_37_1649_2012} evaluates to $v_\text{Q}\comm{\equiv}\langle\langle\ft_\text{c}|\ft_\text{c}\rangle\rangle/\ft_\text{c}^2(\mr_\text{c})a^2 = 1.441 - 0.055\text{i}$ (the size of the contribution from $|x|>x_0$ in this case is $2|I_{x>x_0}|/\ft_\text{c}^2(\mr_\text{c})a^2=0.005$). 
We note, that when the cavity mode leaks into homogeneous surroundings as well as into a guided waveguide mode, as in PC membranes in three dimensions, for example, the integrand in Eq.~(\ref{Eq:SauvanNorm}) will tend to zero in regions of homogeneous material far from the cavity, whereas it will oscillate and increase in magnitude for positions inside the waveguide. Therefore, we expect the above prescription to generalize immediately to such systems as well.

\begin{figure}[htb]
\includegraphics[width=\columnwidth]{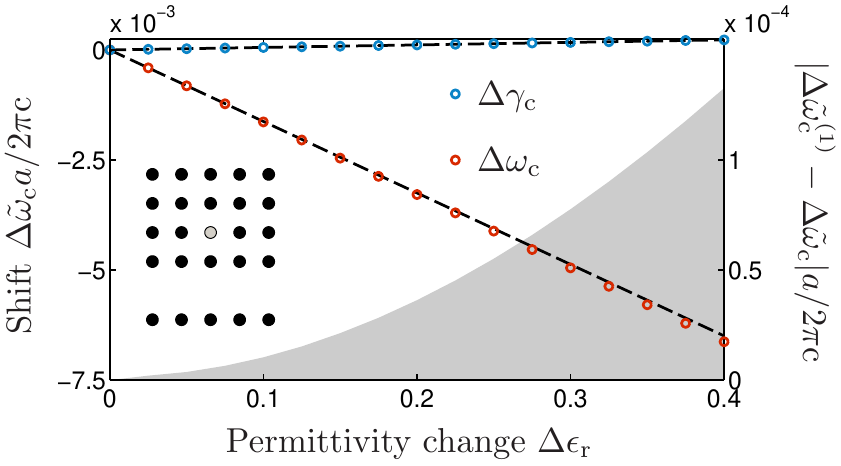}
\caption{\label{Fig:side_coupled_2D_quadratic_rods_oneBlock_dk0_vs_dEps_fancy_scaled}Perturbative change in the resonance frequency \comm{$\Delta\tlo_\text{c} = \Delta\omega_\text{c}-\text{i}\Delta\gamma_\text{c}$} as a function of permittivity change $\Delta\epsilon_\text{r}$ of a cylinder in the center of the cavity as indicated in the inset. \comm{Red and blue circles \comm{(left axis)} show changes in resonance frequency and decay rate, respectively, using full numerical calculations, and dashed black lines show the first order perturbation result of Eq.~(\ref{Eq:perturbationQuasinormalModes}).} Gray shading (right axis) shows the error in the perturbation.}
\end{figure}

Last, as an example application, we use the norm in Eqs.~(\ref{Eq:SauvanNorm}) and (\ref{Eq:divergentGeometric}) to study the influence of small material changes. Using perturbation theory, the first order change in the resonance frequency due to a small change in the permittivity distribution is given as
\begin{align}
\Delta\tlo^\text{\fontsize{2mm}{1em}\selectfont(1)}_\text{c}=-\frac{\tlo_\text{c}}{2\langle\langle\mft_\text{c}|\mft_\text{c}\rangle\rangle}\int_{V}\Delta\epsilon_\text{r}(\mr)\mft_\text{c}(\mr)\cdot\mft_\text{c}(\mr)\ud \mr,
\label{Eq:perturbationQuasinormalModes}
\end{align}
where $\Delta\epsilon(\mr)$ is the position dependent change in permittivity. Eq.~(\ref{Eq:perturbationQuasinormalModes}) is identical to the result for QNMs leaking to homogeneous media~\cite{Lai_PRA_41_5187_1990, Lee_JOSAB_16_1418_1999}; \rev{it may be derived, for example, using the Hellmann-Feynman theorem~\cite{Allen_PRB_33_5611_1986} in an operator formulation of Maxwell's equations similar to Ref.~\cite{Daniel_JOSAB_28_2207_2011} but with a non-Hermitian approach using the QNM norm in Eq.~(\ref{Eq:SauvanNorm}).} 
We consider perturbations to the material system in which an additional low-permittivity rod is inserted in the center of the cavity. As shown in Fig.~\ref{Fig:side_coupled_2D_quadratic_rods_oneBlock_dk0_vs_dEps_fancy_scaled}, the perturbation result of Eq.~(\ref{Eq:perturbationQuasinormalModes}) agrees well with full numerical FEM calculations even for relatively large shifts of several linewidths.

In summary, we have described an approach for numerical calculation of QNMs in cavities coupled to waveguides. The approach relies on a non-local boundary condition to correctly model the radiation condition for light leaking through the waveguide. In addition, we have shown how to normalize these QNMs by use of the theory of divergent series to 
assign a finite value to the divergent integral appearing in the norm. Finally, we have used the normalization procedure for perturbation theory calculations to illustrate how it leads to the correct first order predictions of the resonance shift.

\comm{This work was supported by the Carlsberg Foundation and the Danish Council for Independent Research (FTP 10-093651)}






\end{document}